\documentclass[leqno]{article}

\usepackage[utf8]{inputenc}
\usepackage[english]{babel}
\usepackage[T1]{fontenc}
\usepackage{amsthm, amsmath, amsfonts, amssymb}
\usepackage{url}
\usepackage{hyperref}
\usepackage{authblk}
\usepackage{enumitem}
\usepackage[acronym,nonumberlist,hyperfirst=false]{glossaries}
\usepackage{algorithm}
\usepackage{algorithmicx}
\usepackage{algpseudocode}
\usepackage{complexity}

\setlist[enumerate]{
  wide =0.5\parindent,
  listparindent=0pt
}

\theoremstyle{definition}
\newtheorem{theorem}{Theorem} % [section]
\newtheorem{proposition}{Proposition} % [section]

\newtheorem{cor}{Corollary}

\theoremstyle{definition}
\newtheorem{defn}{Definition}
\newtheorem*{notation*}{Notation}

\newcommand{\set}[1]{\ensuremath{#1}}

\newcommand{\graphFbas}[1]{\ensuremath{G_{#1}}}
\newcommand{\sccRel}[1]{\ensuremath{\leq_{#1}}}

\newcommand{\fbas}{\(\mathsf{FBAS}\)}
\newcommand{\plainFbas}{{\textsf{plain-}\fbas}}
\newcommand{\xyFbas}{\textsf{x-y-}\fbas}
\newcommand{\hep}{\(\mathsf{HEP}\)}
\newcommand{\coNPC}{{\ComplexityFont{coNP \lower-.12em\hbox{\textrm{-}} complete}}}
\newcommand{\w}[1]{\ComplexityFont{W[#1]}}
\newcommand{\whard}[1]{\ComplexityFont{W[#1] \lower-.12em\hbox{\textrm{-}} hard}}
\newcommand{\wh}{\ComplexityFont{W[1] \lower-.12em\hbox{\textrm{-}} hard}}
\newcommand{\wc}[1]{\ComplexityFont{W[#1] \lower-.12em\hbox{\textrm{-}} complete}}
\newcommand{\PC}{\ComplexityFont{P \lower-.12em\hbox{\textrm{-}} complete}}
\renewclass{\NPC}{NP \lower-.12em\hbox{\textrm{-}} complete}
\newcommand{\pSet}[1]{\mathsf{#1}}
\newcommand{\pEl}[1]{\ensuremath{#1}}
\newcommand{\vqp}{\(\mathsf{VQP}\)}
\newclass{\NPH}{NP \lower-.12em\hbox{\textrm{-}} hard}
\newclass{\coNPH}{coNP \lower-.12em\hbox{\textrm{-}} hard}
\setkeys{glslink}{hyper=false}

\newcommand{\mcvp}{\(\mathsf{MCVP}\)}

\newacronym{dqp}{DQP}{Disjoint Quorums Problem}
\newacronym{dqpk}{DQP ($k$)}{Disjoint Quorums Problem ($k$)}
\newacronym{mqp}{MQP}{Minimum Quorum Problem}
\newacronym{mqpk}{MQP (\(k\))}{Minimum Quorum Problem (\(k\))}
\newacronym{mqpkr}{MQP (\(k\), \(r\))}{Minimum Quorum Problem (\(k\), \(r\))}
\newacronym{qsp}{QSP}{Quorum Subset Problem}
\newacronym{xyDqp}{x-y-DQP}{x-y Disjoint Quorums Problem}
\newacronym{xyMqp}{x-y-MQP}{x-y Minimum Quorum Problem}
\newacronym{xyMqpk}{x-y-MQP (\(k\))}{x-y Minimum Quorum Problem (\(k\))}
\newacronym{cliquek}{Clique (\(k\))}{Clique (\(k\))}
\newacronym{xyQsp}{x-y-QSP}{x-y Quorum Subset Problem}
\newacronym{ssp}{SSP}{Set Splitting Problem}
\newacronym{fbas}{\fbas}{Federated Byzantine Agreement System}
\newacronym{plainFbas}{\plainFbas}{Plain Federated Byzantine Agreement System}
\newacronym{xyFbas}{\xyFbas}{x-y Federated Byzantine Agreement System}
\newacronym{fpt}{FPT}{fixed-parameter tractable}
\newacronym{p2p}{p2p}{peer-to-peer}
\newacronym{hep}{\hep}{Horn Entailment Problem}
\newacronym{hsat}{\hep}{Horn Satisfiability Problem}
\newacronym{vqp}{\vqp}{Verify Quorum Problem}
\newacronym{mcvp}{\mcvp}{Monotone Circuit Value Problem}
\newacronym{pbft}{pbft}{Practical Byzantine Fault Tolerance}
\newacronym{scp}{SCP}{Stellar Consensus Protocol}

\title{Complexity of the quorum intersection property of the \acrlong{fbas}}
\author{Łukasz Lachowski \\
  \href{mailto:lukasz.lachowski@tcs.uj.edu.pl}{\texttt{lukasz.lachowski@tcs.uj.edu.pl}}}
\affil{Department of Theoretical Computer Science \\ Faculty of Mathematics and Computer Science \\
  Jagiellonian University \\ Łojasiewicza 6, 30--348 Kraków \\ Poland}
\date{\today}

\begin{document}

\maketitle

\begin{abstract}
  A Federated Byzantine Agreement System (\cite{mazieres2016stellar}) is defined as a pair
  \((\set{V}, \set{Q})\) comprising a set of nodes \(\set{V}\) and a quorum function \(\set{Q}:
    \set{V} \mapsto 2^{2^{\set{V}}} \setminus \{\emptyset\}\) specifying for each node a set of
  subsets of nodes, called quorum slices. A subset of nodes is a quorum if and only if for each of
  its nodes it also contains at least one of its quorum slices. The Disjoint Quorums Problem answers
  the question whether a given instance of \acrlong{fbas} contains two quorums that have no nodes in
  common. We show that this problem is \(\NPC\). We also study the problem of finding a quorum of
  minimal size and show it is \(\NPH\). Further, we consider the problem of checking whether a given
  subset of nodes contains a quorum for some selected node. We show this problem is \(\PC\) and
  describe a method that solves it in linear time with respect to number of nodes and the total size
  of all quorum slices. Moreover, we analyze the complexity of some of these problems using the
  parametrized point of view.
\end{abstract}

\acrlong{fpt}

\section{Introduction}

The problem of reaching consensus among remote process is at the core of many distributed algorithms
like distributed file systems, database management systems and fault tolerant distributed
applications. Researchers studied this problem in many different settings and network models. One of
the well-known form of this problem is the Byzantine fault tolerant state machine replication
problem in asynchronous networks. In this problem one assumes that some subset of nodes might be
unreachable and can even arbitrary lie during a protocol's run. One of the best known variants of
Byzantine agreement is the \acrlong{pbft} protocol~\cite{Castro:1999:PBF:296806.296824}, which was a
major milestone in making Byzantine agreement practical. One disadvantage of this approach is the
closed membership of peers taking part in achieving consensus, making it rather impractical for open
membership network models like the \acrlong{p2p} network model. A solution which tries to tackle
this issue is a protocol called \gls{scp}, presented in~\cite{mazieres2016stellar}. Author
introduces in it a model called \gls{fbas}, of which the \gls{scp} protocol is one possible
realization, and proves its basic properties and correctness. The main goal of that work was to
create a fast, reliable and secure distributed system for financial interactions. The model of
\gls{fbas} consists of a pair \((\set{V}, \set{Q})\), where \(\set{V}\) is a set of nodes and
\(\set{Q}\) is a function, called quorum function, of type \({\set{V} \to 2^{2^{\set{V}}} \setminus
  \{\emptyset\}}\) specifying for each node one or more subsets of nodes, called quorum slices,
where a node belongs to all of its quorum slices. Further, author defines a quorum of an \gls{fbas}
\(F = (\set{V}, \set{Q})\) as a nonempty subset of its nodes \(\set{U} \subseteq \set{V}\) which
contains a quorum slice for each of its members, i.e. \(\forall v \in \set{U} \, \exists q \in
\set{Q}(v) \text{ such that } q \subseteq \set{U}\). For purpose of proving safety and correctness
properties of this model, author uses the assumption that all quorums of a given \gls{fbas} are
intersecting. Therefore, devising an algorithm which can verify this property is crucial. Using the
definition of a quorum we can introduce notions of minimal and minimum quorums. We call a quorum
minimal if none of its proper subsets is also a quorum. A minimum quorum is the smallest cardinality
quorum of a given instance of \gls{fbas}. Notice, that every minimum quorum is minimal, but not all
minimal quorums must also be minimum. In this work we consider computational complexity of several
problems concerning properties of quorums of \gls{fbas}, namely the problem of checking whether
all quorums intersect, problem of finding a minimal quorum and verifying if a given set of nodes
contains a quorum.

We present these problems in two slightly different settings, that is one where all quorum slices
are given explicitly, by some enumeration of their elements, and another where multiple quorum
slices are encoded by definitions of the form ``\(x\) out of \(y\)'', where \(y\) can be some
enumeration of nodes or further definition of this form. Naturally, we assume that every such
definition is finite. We call them \gls{plainFbas} and \gls{xyFbas} respectively. Obviously, this
two different approaches affects the way we measure the size of an instance. Using the ``\(x\) out
of \(y\)'' type of definition we are able to encode exponential number of quorum slices relative to
the amount of space used.
\begin{defn}
  \begin{enumerate}
  \item A {\plainFbas} is an instance of \gls{fbas} \(F = (\set{V}, \set{Q})\) where for every node
    \(v \in \set{V}\) its definition of quorum slices \(\set{Q}(v)\) is given by explicit
    enumeration of subsets of nodes of \(\set{V}\), e.g. \(\set{Q}(v) = \{ \{v, v_1, v_2, v_3\},
    \{v, v_4, v_5, v_6\} \}\). Size of an instance of \plainFbas{} \(F = (\set{V}, \set{Q})\) is equal
    to \( |\set{V}| + \sum_{v \in \set{V}} \sum_{\set{q} \in \set{Q}(v)} |\set{q}|\).
  \item {\xyFbas} is an instance of \gls{fbas} \(F = (\set{V}, \set{Q})\) where for every node \(v
    \in \set{V}\) its corresponding definition of quorum slices \(\set{Q}(v)\) is given by
    enumeration of declarations of the form ``\(x\) elements of \(\{\dots\}\)'', e.g.
    \begin{flalign*}
    \set{Q}(v) =
    & \{ \text{two elements of } \{v_1, v_2, v_3\}, & & \\
    & \text{one element of } \\
    & \quad \{ \text{one element of } \{v_4, v_5\}, & \\
    & \quad \text{two elements of } \{v_6, v_7, v_8\} \} \}
    \end{flalign*}
    Size of an instance of
    {\xyFbas} is is equal to \(|V| + \sum_{v \in \set{V}} \sum_{(x \text{ elements of } \set{q}) \in
      \set{Q}(v)} |\set{q}|\). Notice that notion of size of \(\set{q}\) is defined recursively,
    depending on whether it is an enumeration of nodes or further definition of the form ``x of y''.
    In case \(\set{q}\) is of the form \(\{v_1, v_2, \ldots, v_n\}\), its size \(|\set{q}|\) is
    equal to its cardinality. Otherwise, if it is a collection of definitions of the form ``x of
    y'', we define its size recursively as the sum of sizes of its elements.
  \end{enumerate}
\end{defn}
Problems which we study in this work are formally defined below.
\begin{description}
\item[\gls{dqp}]
  \hfill\\\textit{Instance:} An instance of \gls{plainFbas}.
  \hfill\\\textit{Goal:} Determine if the given instance of \gls{plainFbas} has two disjoint
  quorums.
\item[\gls{mqpk}]
  \hfill\\\textit{Instance:} An instance of \gls{plainFbas} and an nonnegative integer \(k\).
  \hfill\\\textit{Goal:} Determine if the given instance contains a quorum of size \(k\).
\item[\gls{qsp}]
  \hfill\\\textit{Instance:} An instance \((\set{V}, \set{Q})\) of \gls{plainFbas} together with
  a node \(v \in \set{V}\) and a subset of nodes \(\set{q} \subseteq \set{V}\).
  \hfill\\\textit{Goal:} Determine if the subset \(\set{q}\) contains a quorum that includes node
  \(v\).
\end{description}
We also consider each of these problems assuming that the given instance of \gls{fbas} is declared
using the ``\(x\) out of \(y\)'' type of definition. We call these versions \gls{xyDqp},
\gls{xyMqpk} and \gls{xyQsp} respectively.

We study complexity aspects of these problems both from classical and parameterized point of view
(\cite{flum2006parameterized}). The remainder of this work is organized as follows. In Section
\ref{section:dqp} we consider the problem of searching for disjoint quorums and prove that it is
\(\NPC\). Since this problem does not depend on any numerical value we can conclude it is strong
\(\NPC\). Further, we show that it remains hard for a very restricted family of \gls{fbas}
instances, where each node can have at most two quorum slices and every quorum slice has at most two
elements. We conclude this part providing characterization of family of \gls{fbas} instances for
which the \gls{dqp} can be solved in polynomial time. Next, we show that problem of finding minimum
quorum is \(\NPH\) and that it remains hard for same family of \gls{fbas} instances as we used in
previous result. Further, we focus on solving the \acrlong{qsp}. We present a solution that runs in
linear time and also prove that it is \(\PC\). In Section~\ref{section:fpt} we describe the
parameterized complexity of some of these problems. We try to find parameterizations of \gls{dqp}
and present a simple randomized algorithm solving it. We also show that \gls{xyMqpk} is
\(\whard{1}\). The problem of finding an \acrshort{fpt} algorithm for \gls{mqpk} remains open.

\section{Hardness results}\label{section:dqp}

The first problem which we consider in this section is the complexity of verifying whether a given
instance of \gls{fbas} contains two disjoint quorums. We show that this problem is \(\NPC\) and try
to limit the type of configurations for which it remains hard. We also provide a characterization of
family of \gls{fbas} configurations that seems to be consistent with the intuition of how new nodes
should be declared when they are attached to some already existing instance of \gls{fbas}. Moreover,
the way we define this family of configurations ensures that all of its quorums are intersecting.
Membership for that class of configurations can be verified in polynomial time. Next, we analyze the
problem of finding a quorum of minimal cardinality and prove that it is \(\NPH\). Lastly, we provide
an efficient algorithm for the \acrlong{qsp} and show that it is \(\PC\).

\begin{theorem}\label{theorem:dqp-npc}
  The \acrlong{dqp} is \NPC{}.
  \begin{proof}
    \gls{dqp} is in \(\NP\), since we can verify in polynomial time if a certificate consisting of
    two subsets of nodes defines two separate quorums. For proving it is \(\NPH\) we present a
    reduction from the \gls{ssp}, which is one of Garey\&Johnson's classical \(\NPC\)
    problems~\cite{garey1979computers}. The input of the \gls{ssp} consist of a family \(\pSet{F}\)
    of subsets of a finite set \(\pSet{S}\). A solution to this problem should answer \(yes\) if and
    only if it is possible to partition the set \(\pSet{S}\) into two subsets \(\pSet{S_1}\),
    \(\pSet{S_2}\), such that all elements of \(\pSet{F}\) are split by this partition. Given an
    instance of this problem, our reduction builds in polynomial time an instance of \gls{dqp}, such
    that the answer for the \gls{ssp} is \(yes\) if and only if the corresponding answer to
    \gls{dqp} is also \(yes\). Our reduction constructs an instance of \gls{dqp} as follows. First,
    we create a new node of \acrshort{fbas} for every element \(\pEl{x}\) of the given set
    \(\pSet{S}\). For each \(\pEl{x}\) we denote such node by \(\pEl{v_x}\). For every subset
    \(\pEl{f} \in \pSet{F}\) we create multiple instances of \acrshort{fbas} nodes, one for each
    element of the set \(\pSet{S}\). We denote these nodes by \(\pEl{v^{f}_{x}}\). Every node
    \(\pEl{v_x}\) defines exactly one quorum slice, which consists of nodes indexed by the same
    corresponding element \(\pEl{x}\), i.e. \(\pEl{v^{f}_{x}}\). Every node \(\pEl{v^{f}_{x}}\),
    which corresponds to a subset \(\pEl{f}\), defines a separate quorum slice for each element
    \(\pEl{y} \in \pEl{f}\) of this subset. Such slice points at the corresponding node for the
    element \(\pEl{y}\) in \acrshort{fbas}, i.e. \(\pEl{v_y}\). Now we prove that this is a valid
    polynomial time reduction. Obviously, our reduction is computable in polynomial time with
    respect to the size of the \acrshort{ssp} instance. Let's assume we can partition the set
    \(\pSet{S}\) into two subsets \(\pSet{S_1}\) and \(\pSet{S_2}\) and that each element of the set
    \(\pSet{F}\) is split by these sets. We argue that we can find two separate quorums using this
    partition. Each quorum consists of all corresponding nodes of \acrshort{fbas}, i.e. if \(\pEl{x}
    \in \pSet{S_i}\), then \(\pEl{v_x} \in \pSet{Q_i}\), for \(i\) equal \(1\) or \(2\). As every
    such element requires a quorum slice consisting of all nodes that are denoted by
    \(\pEl{v^{f}_{x}}\), we also need to include them in our quorums. Set of these nodes is disjoint
    for every node \(\pEl{v_x}\). Each of these nodes corresponds to one of the elements of
    \(\pSet{F}\) and we know we can split each of these sets by \(\pSet{S_1}\) and \(\pSet{S_2}\).
    Hence, we can find at least one element which will be part of the same quorum as the
    corresponding node \(\pEl{v_x}\) and fulfill requirements of one of the quorum slices for every
    node \(\pEl{v^{f}_{x}}\). We see that we get this way two separate quorums. Now assume we can
    find two separate quorums in our reduced instance. Since we have two types of nodes, denoted by
    \(\pEl{v_x}\) and \(\pEl{v^{f}_{x}}\) respectively, and each of the nodes denoted by
    \(\pEl{v^{f}_{x}}\) requires at least one node \(\pEl{v_y}\) that corresponds to some element
    \(\pEl{y}\) contained in the set \(\pEl{f}\), we can conclude that every quorum must consist of
    some node that is denoted by \(\pEl{v_y}\). We can ignore all nodes that corresponds to some
    subset \(\pEl{f}\) and are denoted by \(\pEl{v^{f}_{x}}\) for building our splitting sets, cause
    they do not correspond to elements of the set \(\pSet{S}\). We select one of the quorums and
    choose one of the elements \(\pEl{v_x}\) from it. We put the associated element \(\pEl{x}\) into
    the splitting set \(\pSet{S_1}\). The way we constructed its quorum slice requires to pick at
    least one quorum slice for each of its corresponding nodes \(\pEl{v^{f}_{x}}\). For every subset
    \(\pEl{f}\) each of these nodes requires at least one of the elements associated with the subset
    \(\pSet{f}\). We select one such node \(\pEl{v_y}\) for each of the nodes \(\pEl{v^{f}_{x}}\)
    and put its associated element from the set \(\pSet{S}\) into \(\pSet{S_1}\). This can be
    translated as choosing at least one element for each member of the set \(\pSet{F}\), which
    proves that we can properly split our set \(\pSet{S}\).
    % We can simply add such element to our set \(\pSet{S_1}\).
    As we assumed that we have two separate quorums, we can be sure that one of the nodes
    corresponding to elements of the subset \(\pEl{f}\) belongs to that other quorum. This is due
    the fact that the other quorum consist of some element \(\pEl{v_y}\), among others, that
    requires a node \(\pEl{v^{f}_{y}}\), which also requires one of the nodes corresponding to
    elements of the subset \(\pEl{f}\). Following these construction, we can select at least one
    element from each element of set \(\pSet{F}\) and at the same time be sure there is at least one
    element which we did not select from each of these subsets. From this, we see that if we put all
    remaining elements into the set \(\pSet{S_2}\), we receive a correct splitting of the set
    \(\pSet{S}\).
  \end{proof}
\end{theorem}
As an immediate consequence of this theorem we obtain the following result.
\begin{cor}
  \gls{xyDqp} is \NPC.
\end{cor}
We can derive a stronger result restricting the class of configurations for which this problem
remains \(\NPC\).
\begin{cor}\label{cor:smallQ_dqp}
  \gls{dqp} remains \(\NPC\) for \gls{fbas} configurations in which each node has at most two quorum
  slices and each quorum slice contains at most two nodes.
  \begin{proof}
    First, we show that every node declared in the proof of Theorem~\ref{theorem:dqp-npc} that has
    more than two quorum slices can be remapped onto polynomial number of new nodes, each having at
    most two quorum slices. Consider a node \(v\) that has exectly three distinct quorum slices,
    i.e. \(\set{Q}(v)\) equals \(\{q_1, q_2, q_3\}\). We can redefine its set of quorum slices
    \(\set{Q}(v)\) onto \(\set{Q}(v) = \{q_1, \{v'\}\}\), where \(v'\) is a new node such that
    \(\set{Q}(v') = \{q_2, q_3\}\). Similarly, if for some node \(v\) one of its quorum slices has
    exactly three nodes, that is for some quorum slice \(q\) it is equal to \(\{v_1, v_2, v_3\} \in
    \set{Q}(v)\), then we can redefine such \(q\) onto \(q' = \{v_1, v'\}\), where \(v'\) is a new
    node for which \(\set{Q}(v')\) is equal to \(\{\{v_2, v_3\}\}\). We observe that for such
    remapped instance the answer of \gls{dqp} is \(yes\) if and only if the answer for the initial
    instance of \gls{dqp} is \(yes\). The size of this new instance is no bigger than twice the size
    of the original instance.
  \end{proof}
\end{cor}
We can also easily derive some result regarding the problem of finding quorums that intersect with
some specified number of nodes.
\begin{cor}
  The problem of finding two quorums that share at most \(k\) nodes is \(\NPC\).
\end{cor}
Now, we try to find some useful properties of minimal quorums which we can exploit in order to find
disjoint quorums. For an instance of \gls{fbas} \(F = (\set{V}, \set{Q})\), let \(\graphFbas{F}\)
denote a representation of \(F\) by a directed graph. We build such representation by creating a
vertex for every node of \(F\) and by setting a directed edge between vertices \(a\) and \(b\) if
and only if the corresponding node for vertex \(b\) is contained within one of the quorum slices of
node mapped from the vertex \(a\). We can observer that every minimal quorum is a strongly connected
component of such graph representation.
\begin{proposition}\label{lemma:mq_scc}
  Every minimal quorum \(\set{q}\) of an \gls{fbas} \(F = (\set{V}, \set{Q})\) is a strongly
  connected component of its graph representation \(\graphFbas{F}\), that is for a minimal quorum
  \(\set{q} \subseteq \set{V}\) its corresponding induced subgraph \(\graphFbas{F}[\set{q}]\) is a
  strongly connected component.
  \begin{proof}
    It easy to notice that for two nodes \(v_1, v_2 \in \set{q}\), such that \(v_2\) is not
    reachable from \(v_1\) in \(\graphFbas{F}[\set{q}]\), a set of all nodes reachable from \(v_1\)
    in \(\graphFbas{F}[\set{q}]\) contains a quorum \(\set{q'} \subsetneq \set{q}\).
  \end{proof}
\end{proposition}
For an instance of \gls{fbas} \(F\), by \sccRel{F} we denote the relation defined between strongly
connected components of \graphFbas{F} that corresponds with the reachability relation of such
components. Using Proposition~\ref{lemma:mq_scc} we can derive the following result.
\begin{cor}\label{cor:scc_minimum}
  If for a given instance of \gls{fbas} \(F = (\set{V}, \set{Q})\) all of its quorums are pairwise
  intersecting, then its corresponding relation \(\sccRel{F}\) has the greatest element (some
  strongly connected component of \graphFbas{F}) that contains all minimal quorums.
\end{cor}
Our next goal is to introduce some method for enumerating all quorums. A~simple way to achieve this
goal would be a procedure that enumerates all subsets of a given set of nodes and verifies if such
subset is a quorum. Unfortunately, this method requires time proportional to \(2^n{n}^{O(1)}\)
regardless of number of quorums of a given instance of \gls{fbas}, which makes it impractical for
even small values of \(n\). Instead, we can devise a method that while it is enumerating such
candidates simply filters out subsets that does not contain a quorum.
\begin{algorithm}[th]
  \caption{Disjoint Quorums}\label{algo:algorithm_scc}
  \hspace*{\algorithmicindent} \textbf{Input}: An instance of \gls{fbas} $F = (\set{V}, \set{Q})$. \\
  \hspace*{\algorithmicindent} \textbf{Output}: \textit{YES} if there exists two disjoint quorums
  $q_!, q_2 \subseteq \set{V}$ of $F$, or \textit{NO} if all quorums of $F$ intersect.
  \begin{algorithmic}[1]
    \State $sccs \gets \text{ set of strongly connected components of } F$
    \State $sccsTop \gets \text{ topological order of } sccs$
    \State $top \gets \text{some greatest element of the topological order } sccsTop$
   \ForAll {$q \gets sccs \setminus \{top\}$}
      \If {$q$ contains a quorum}
      \State\Return \textit{YES}
      \EndIf
    \EndFor
    \ForAll{$q \gets \Call{Quorums}{top}$}
      \If {$top \setminus \{q\}$ contains a quorum}
        \State\Return \textit{YES}
      \EndIf
    \EndFor
    \State\Return \textit{NO}
  \end{algorithmic}
\end{algorithm}
\begin{proposition}
  Quorums of an \gls{fbas} can be enumerated with polynomial-time delay.
  \begin{proof}
    We use classical branching strategy. For a given instance of \gls{fbas} \(F = (\set{V},
    \set{Q})\) let \(\sigma = (v_1, v_2, \ldots, v_n)\) be an arbitrary ordering of nodes
    \(\set{V}\). We start with two sets \(\set{V_1} = \set{V}\) and \(\set{V_2} = \emptyset\). The
    set \(\set{V_1}\) contains nodes which we are going to consider and the set \(\set{V_2}\)
    represents nodes we require to be part of some quorum. Then, we consider vertices in order
    \(\sigma\) and we branch into two subcases: a node \(v_i\) is removed from the set \(\set{V_1}\)
    or \(v_i\) is removed from the set \(\set{V_1}\) and put into the set \(\set{V_2}\). The moment
    we discover that the set \(\set{V_2}\) is a quorum we output it and continue our strategy. We
    terminate a branch if we discover the set \(\set{V_2}\) is not contained within any quorum
    contained in the set \(\set{V_1} \cup \set{V_2}\). We can verify it by using a simple iterative
    fix-point strategy that filters out nodes that do not have any quorum slice within a given set
    of nodes, i.e. we search for a fix-point of the function \(f: \set{V} \to \set{V}\) given by
    \(f(x) = \{v: v \in x \text{ and } v \text{ has a quorum slice in } x \}\). Since every quorum
    is a fixed-point of \(f\) and \(f\) is a monotonic function defined on a complete lattice, we
    see, using the Kleene Fixed-Point Theorem, that if we start our iteration using the set
    \(\set{V_1} \cup \set{V_2}\) we obtain a fixed-point of \(f\) that is the maximal quorum
    contained within this set. Using this method we only consider branches that produce some unique
    quorum and each quorum is returned within polynomial waiting time, proportional to the depth of
    the branching tree times the time required for computations at each node of the branching tree.
  \end{proof}
\end{proposition}
Using this enumeration procedure together with Corollary~\ref{cor:scc_minimum}, we can define a
simple algorithm solving \gls{dqp}. Our method is displayed as Algorithm~\ref{algo:algorithm_scc}.
Based on this algorithm we can also think of some guideline for defining node's quorum slices that
ensures the quorum intersection property and that is also easy to verify. To this point, one way to
achieve this goal is to configure each node using the following rules:
\begin{itemize}
\item there should be exactly one strongly connected component of \graphFbas{F} that is the greatest
  element of the relation \sccRel{F}
\item declaration of node's \(v\) quorum slices \(\set{Q}(v)\) should be of the form: more than half
  of the nodes from its strongly connected component in \graphFbas{F} and any node from some other
  strongly connected component of \graphFbas{F}
\end{itemize}
It is easy to see that if an instance of \gls{fbas} is defined using these rules then all of its
quorums are intersecting and that verifying such properties can be achieved in polynomial time.

Now we consider the problem of finding some smallest cardinality quorum. We show that
this problem is \(\NPH\) and that it remains hard for the class of configurations where each
node has at most two quorum slices and each quorum slice contains at most two nodes.
\begin{theorem}\label{theorem:mqp}
  The Minimal Quorum Problem is \(\NPH\).
  \begin{proof}
    The proof uses reduction from the Vertex Cover Problem, shown to be \(\NPH\) by Karp in
    \cite{Karp1972}. For a given graph \(G = (\set{V}, \set{E})\) we construct an instance of \gls{fbas}
    as follows. We create a new node of \gls{fbas} for every vertex and edge of the graph. For each
    node that was mapped from some edge of the given graph, we create two quorum slices, each
    pointing at a node corresponding to one of the endpoints of that edge. For every node
    corresponding to some of the graph's vertices, we create a single quorum slice containing all of
    the nodes mapped from the graph's edges. It is easy to prove that a given instance of the Vertex
    Cover Problem has a solution of size \(k\) if and only if the corresponding instance of
    \gls{fbas} has a quorum of size \(|E| + k\).
  \end{proof}
\end{theorem}
We can use similar technique as we used in Corollary~\ref{cor:smallQ_dqp} to prove slightly stronger
result.
\begin{cor}\label{cor:mqp}
  \gls{mqp} remains \(\NPH\) for configurations in which each node has at most two quorum slices and
  each quorum slice contains at most two nodes.
\end{cor}
\begin{cor}
  \gls{xyMqp} is \(\NPH\).
\end{cor}

To conclude this section, we show that the \acrlong{qsp} is \(\PC\) and devise an algorithm that
solves it in linear time.
\begin{theorem}
  The \acrlong{qsp} is \(\PC\). Moreover, it can be solved in linear time with respect to the total size
  of all quorum slices.
  \begin{proof}
    We present a log space reduction from the \gls{mcvp}, which was proved \(\PC\) by Goldschlager
    in \cite{Goldschlager:1977:MPC:1008354.1008356}. The input of the \gls{mcvp} is composed of a
    set of logical gates \(\pEl{g_1}, \pEl{g_2}, \dots, \pEl{g_n}\) where each is an \(and\) gate
    \(\pEl{g_i} = \pEl{g_j} \land \pEl{g_k}\), an \(or\) gate \(\pEl{g_i} = \pEl{g_j} \lor
    \pEl{g_k}\) (\(j\) and \(k\) are smaller than \(i\)) or a constant value \(\pEl{g_i} = true\) or
    \(\pEl{g_i} = false\). Using these definitions we wish to compute the value of the gate
    \(\pEl{g_n}\). Our reduction outputs a new element of the set \(\set{V}\) in the definition of
    \acrshort{fbas} for every declared gate. Next, we define the set \(\set{W}\) and all quorum
    slices for all created nodes. Our reduction transforms each \(and\) gate \(\pEl{g_i}=\pEl{g_j}
    \land \pEl{g_k}\) by including its corresponding element of \acrshort{fbas} in the definition of
    the set \(\pSet{W}\) and outputting a single quorum slice for it of the form \(\{\pEl{g_i},
    \pEl{g_j}, \pEl{g_k}\}\). Similarly, we map each \(or\) gate \(\pEl{g_i}=\pEl{g_j} \lor
    \pEl{g_k}\) into a new element of the set \(\pSet{W}\) and two of its quorum slices of the form
    \(\{\pEl{g_i}, \pEl{g_j}\}\) and \(\{\pEl{g_i}, \pEl{g_k}\}\). For all gates defined as
    \(\pEl{g_i}=true\) we create an element of the set \(\set{W}\) together with a quorum slice
    containing just the corresponding node \(\{\pEl{g_i}\}\). For all \(false\) gates
    \(\pEl{g_i}=false\) we create a single quorum slice containing the corresponding element of
    \gls{fbas} as the only element and do not modify the set \(\set{W}\). Obviously, this a log
    space reduction. We now prove that a solution for the \gls{qsp} answers \(yes\) for a given
    instance if and only if the mapped node of the gate \(\pEl{g_n}\) has a quorum in our reduced
    instance of \gls{qsp}. We will prove it by induction on the size of the circuit. We obtain the
    base case of the induction immediately, since for each constant gate with value \(true\) we put
    all elements of its only quorum slice into the set \(\set{W}\) for a corresponding node of
    \acrshort{fbas}. In case of a \(false\) constant we see that its only quorum slice is not
    included in the set \(\set{W}\) and so this instance has no quorum. Now assume we have a list of
    gates \(\{\pEl{g_1}, \pEl{g_2}, \dots, \pEl{g_n}\}\). In case the output gate is of \(and\)
    type, namely \(\pEl{g_n}=\pEl{g_j} \land \pEl{g_k}\), we see it is evaluated to \(true\) if and
    only if both \(\pEl{g_j}\) and \(\pEl{g_k}\) are also evaluated to \(true\). Using the inductive
    hypothesis, we see that this is the case if and only if their corresponding nodes have quorums
    in our reduced instance. In case of an \(and\) gate our reduction adds a new node and a single
    quorum slice containing corresponding nodes for \(\pEl{g_j}\) and \(\pEl{g_k}\). Since we also
    add a node corresponding to gate \(\pEl{g_n}\) to the set \(\set{W}\), we see that we can find a
    quorum for it. Similarly, if the gate \(\pEl{g_n}\) is an \(or\) gate, namely
    \(\pEl{g_n}=\pEl{g_j} \lor \pEl{g_k}\), then the circuit is evaluated to \(true\) if and only if
    at least one of the gates \(\pEl{g_j}\) or \(\pEl{g_k}\) is evaluated to \(true\). Without loss
    of generality, we can assume this is gate \(\pEl{g_j}\). Using the inductive hypothesis we can
    deduce that \(\pEl{g_j}\) is evaluated to \(true\) if and only if the corresponding node of
    \gls{fbas} has a quorum. Following the reduction for the \(or\) gate, we deduce that gate
    \(g_n\) is evaluated to \(true\) if and only if its corresponding node has a quorum in the set
    \(\set{W}\).

    Now, we argue that \gls{qsp} can be solved in linear time. The main idea of behind our method is
    to exploit more carefully the fix-point strategy described in the proof of
    Theorem~\ref{theorem:dqp-npc}. At each step of this algorithm we need to remove all nodes not
    containing a quorum slice in currently processed subset of nodes. We observer that it can be
    realized efficiently by simple bookkeeping and by storing at every node a list of references to
    all quorum slices containing it. Then, after we remove a node from currently considered subset,
    we can follow this links to build a list of all nodes that needs to be removed in next iteration
    of the fix-point strategy. Since the overall number of links between nodes and quorum slices
    containing them equals the total size of all quorum slices and every such link is visited at
    most once, we conclude that this method requires time linear in the size of a given instance.
  \end{proof}
\end{theorem}
Using similar technique we can devise an algorithm solving \gls{xyQsp} in linear time.
\begin{cor}
  \gls{xyQsp} is \(\PC\) and can be solved in linear time.
\end{cor}

\section{Parameterized complexity results}\label{section:fpt}

The Parametrized Complexity Theory was proposed by Downey and Fellows in
\cite{flum2006parameterized}. It was presented as a promising alternative to deal with \(\NPH\)
problems. The general form of problems considered by this theory is as follows: given an object
\(x\) and a nonnegative integer \(k\), does \(x\) have some property that depends only on \(k\)? For
problems studied by the parametrized complexity theory, \(k\) is fixed as the \textit{parameter} and
assumed to be small in comparison with size of \(x\). For many problems, finding a deterministic
parametrized algorithm whose running time is exponential with respect to \(k\), but polynomial with
respect to size of \(x\), might be very desirable, as it provides an efficient method of
deterministically solving such problems when the parameter \(k\) is reasonably small.

\begin{defn}
  A problem \(A\) is \gls{fpt}, if for any instance \(x\) and parameter \(k\) it can be decided
  whether \((x, k)\) is a yes-instance of \(A\) in running time \(f(k)|x|^{O(1)}\), where \(f\) is
  an arbitrary computable function on nonnegative integers. The corresponding complexity class is
  called \gls{fpt}.
\end{defn}
We also use the notion of \gls{fpt}-reducability between problems, which is defined as follows.
\begin{defn}
  Let \(A\) and \(B\) be two parametrized problems over alphabets \(\Sigma\) and
  \(\Sigma'\), respectively. We define an \gls{fpt}-reduction from \(A\) to \(B\) as a
  mapping \(R: \Sigma \to \Sigma'\), that given an instance \((x, k)\) of \(A\) outputs an instance
  \((x', k')\) of \(B\) such that:
  \begin{enumerate}
  \item \((x, k)\) is a yes-instance of \(A\) if and only if \((x', k')\) is a yes-instance of \(B\),
  \item \(k' \leq g(k)\) for some computable function \(g\),
  \item \(R\) is computable by an \gls{fpt} algorithm with respect to \(k\).
  \end{enumerate}
\end{defn}
Additionally to the \gls{fpt} class, parametrized problems are organized in a hierarchy, called
\(W\)-hierarchy (\(\gls{fpt} = \w{0} \subseteq \w{1} \subseteq \w{2} \subseteq \cdots \subseteq \w{P}\),
that groups them according to their parametrized intractability level. It is conjectured that each
of the containments is proper (\cite{flum2006parameterized}). It is also known that if \(P = NP\)
then this hierarchy collapses (\cite{flum2006parameterized}). For each element from these hierarchy
we also define a notion of \(\w{k}\)-hardness and completeness as follows: a problem \(A\) is
\(\whard{k}\) under \gls{fpt}-reductions if every problem in \(\w{k}\) is \gls{fpt}-reducible to
\(A\); a problem \(A\) is \(\wc{k}\) under \gls{fpt}-reductions if \(A \in \w{k}\) and \(A\) is
\(\whard{k}\).

Our next goal is to find some parameterizations and \gls{fpt} algorithms for \gls{dqp} and \gls{mqp}
problems. In Section \ref{section:dqp} we provided a solution for \gls{qsp}, that is verifying if a
given set of nodes contains a quorum. Presented algorithm runs in polynomial time, regardless of
whether quorum slices are encoded using explicit enumeration of their elements or by using the ``x
of y'' type of definitions. We can use this solution to find a minimal quorum of a given subset of
nodes, by simply trying to greedily remove nodes from it and checking whether such reduced subset
still contains a quorum. We output a subset when we realize that there is no quorum after removal of
any of its nodes. As a result of this procedure we obtain some minimal quorum. If we are lucky, it
can be also of minimal size, but this procedure does not guarantee it. As we showed in Section
\ref{section:dqp}, problem of finding a quorum of minimum size is more complicated.

A possible certificate for \gls{dqp} can simply consist of some enumeration of two disjoint
subsets, each being a quorum. We can parameterize this problem by the size of such certificate, that
is we define the problem \gls{dqpk} as of finding two disjoint quorums for which their combined size
is equal to \(k\). Now, we present an easy probabilistic algorithm based on the technique of random
seperation that solves this problem with constant probability and with time complexity of order
\(O(2^{k}n^{O(1)})\). Our algorithm consists of two subroutines: first that randomly colors each
node by one of two colors, lets say red and green, and second that verifies if each subset induced
by each color contains some quorum.
\begin{proposition}
  Let \(F = (\set{V}, \set{Q})\) be an instance of \gls{fbas} for which size of the set \(\set{V}\)
  equals \(n\), and let \(\set{Q_1}\) and \(\set{Q_2}\) be two disjoint subsets of nodes of combined
  size equal to \(|\set{Q_q}| + |\set{Q_2}| = k\). Let \(\set{X}: \set{V} \to \{\text{red,
    green}\}\) be a coloring of nodes, chosen uniformly at random (i.e. each element of \(\set{V}\)
  is colored with one of two colors uniformly and independently at random). Then the probability
  that elements of \(\set{Q_1}\) are all colored red and elements of \(\set{Q_2}\) are all colored
  green is at least \(2^{-k}\).
  \begin{proof}
    There are \(2^{-k}\) different colorings of sets \(\set{Q_1}\) and \(\set{Q_2}\), and exactly
    one of them colors all nodes of \(\set{Q_1}\) by red and of \(\set{Q_2}\) by green.
  \end{proof}
\end{proposition}
Hence, by our color coding procedure we are able to distinguish two disjoint quorums with
probability at least \(2^{-k}\). Repeating this procedure \(2^{k}\) times gives a constant
probability of success. Combined with the linear-time solution for the \acrlong{qsp} we obtain a
method that solves the \acrlong{dqpk} with constant probability in time proportional to \(2^{k}n\).
Further, we can derandomize this algorithm using the method of constructing \((n,k)\)-universal
sets, see \cite{Naor:1995:SND:795662.796315} for details.

Next, we show that the \gls{xyMqpk} is \(\whard{1}\). To this end, we show an \gls{fpt}-reduction of
the \acrshort{cliquek} problem, that is a problem of finding a clique of size at least \(k\) in a
given graph.
\begin{theorem}
  \acrshort{cliquek} is \gls{fpt}-reducable to \gls{xyMqpk}.
  \begin{proof}
    Given an instance \(G = ((V, E), k)\) of \acrshort{cliquek}, we construct an instance of
    \gls{xyMqpk} as follows. We create a node of \gls{fbas} for every vertex of \(G\). We also
    create some additional node of \gls{fbas} denoted by \(s\). If the degree of a vertex \(v
    \in G\) is smaller than \(k-1\), we create a single quorum slice for its associated node in
    \gls{fbas} requiring only node \(s\). Otherwise, if a vertex \(v \in G\) has \(k-1\) or more
    neighbors, we create a single quorum slice for its associated node of \gls{fbas} of the form
    \(\{ k-1 \text{ of nodes corresponding to \(v's\) neighbors}\}\). Lastly, we create a quorum
    slice for the node \(s\) that is of the form \(\{ \text{all nodes} \}\). It is clear that
    using this reduction, graph \(G\) has a clique of size \(k\) if and only if the corresponding
    instance of \gls{fbas} has a quorum of size \(k\). Hence, \acrshort{cliquek} is \gls{fpt}-reducable
    to \gls{xyMqpk}.
  \end{proof}
\end{theorem}
\begin{cor}
  \gls{xyMqpk} is \(\wh\).
\end{cor}
By Corollary~\ref{theorem:mqp}, \gls{mqp} remains \(\NPH\) for the class of \gls{fbas}
configurations, where each node has at most two distinct quorum slices and every quorum slice has at
most two members. Let \gls{mqpkr} stands for the parametrized version of \gls{mqpk} where every node
of \gls{fbas} has at most \(r\) quorum slices with the same number of nodes. Our next goal is to
show that this problem is in \gls{fpt} for parameters \(k\) and \(r\).
\begin{theorem}
  \gls{mqpkr} can be solved in time proportional to \((kr)^kn^{O(1)}\), hence it is in \gls{fpt}.
  \begin{proof}
    We show that \gls{mqpkr} can be solved in \gls{fpt} time using the bounded search technique.
    Given an instance of \gls{fbas} \(F = (\set{V}, \set{Q})\), we invoke the following search
    method for every node \(v \in \set{V}\). During the execution of the algorithm we keep a set of
    nodes \(\set{W}\) that contains all nodes selected in particular branch. At every step of the
    algorithm we pick a node \(v\) from the set \(\set{W}\) that has no quorum slice in set
    \(\set{W}\) and recursively branch by choosing one of its quorum slices no bigger than \(k\) and
    including it in the set \(\set{W}\). We stop the algorithm whenever the set \(\set{W}\) contains
    more than \(k\) elements or we realize it is a quorum. Clearly, no branch is longer than \(k\)
    and there is at most \(kr\) quorum slices to consider at each node. Therefore, our algorithm
    requires time bounded by \((kr)^kn^{O(1)}\) to process all nodes.
  \end{proof}
\end{theorem}

\section{Conclusions}

In this work we have proved that the \acrlong{dqp} is \(\NPC\). It remains \(\NPC\) for simple class
of configurations where each node has at most two quorum slices and each quorum slice has at most
two elements. We also have showed that the problem of finding a minimal quorum is \(\NPH\) for same
class of configurations. Additionally, we have proved that the \acrlong{qsp} is \(\PC\) and can be
solved efficiently. We have also studied these problems in the context of the theory of fixed
parameter tractability. To this end, we have proved that the \acrlong{xyMqpk} is \(\whard{1}\) and
that it is in \gls{fpt} assuming that each node declares no more than \(r\) quorum slices of each
size. The question of classifying the parametrized problem of \gls{mqpk} and also of finding quorums
that intersect with at most \(k\) nodes remains open.

% \printbibliography
\bibliographystyle{alpha}
\bibliography{quorum_intersection}

\begin{thebibliography}{NSS95}

\bibitem[CL99]{Castro:1999:PBF:296806.296824}
Miguel Castro and Barbara Liskov.
\newblock Practical byzantine fault tolerance.
\newblock In {\em Proceedings of the Third Symposium on Operating Systems
  Design and Implementation}, OSDI '99, pages 173--186, Berkeley, CA, USA,
  1999. USENIX Association.

\bibitem[FG06]{flum2006parameterized}
J{\"o}rg Flum and Martin Grohe.
\newblock {\em Parameterized complexity theory}.
\newblock Springer Science \& Business Media, 2006.

\bibitem[GJ79]{garey1979computers}
Michael~R Garey and David~S Johnson.
\newblock {\em Computers and intractability: A Guide to the Theory of
  \(NP\)-Completeness}.
\newblock 1979.

\bibitem[Gol77]{Goldschlager:1977:MPC:1008354.1008356}
Leslie~M. Goldschlager.
\newblock The monotone and planar circuit value problems are log space complete
  for p.
\newblock {\em SIGACT News}, 9(2):25--29, July 1977.

\bibitem[Kar72]{Karp1972}
Richard~M. Karp.
\newblock {\em Reducibility among Combinatorial Problems}, pages 85--103.
\newblock Springer US, Boston, MA, 1972.

\bibitem[Maz16]{mazieres2016stellar}
David Mazieres.
\newblock The stellar consensus protocol: a federated model for internet-level
  consensus.
\newblock \url{https://www.stellar.org/papers/stellar-consensus-protocol.pdf},
  2016.

\bibitem[NSS95]{Naor:1995:SND:795662.796315}
M.~Naor, L.~J. Schulman, and A.~Srinivasan.
\newblock Splitters and near-optimal derandomization.
\newblock In {\em Proceedings of the 36th Annual Symposium on Foundations of
  Computer Science}, FOCS '95, pages 182--, Washington, DC, USA, 1995. IEEE
  Computer Society.

\end{thebibliography}

\end{document}